\def\be{\begin{equation}}
\def\ee{\end{equation}}
\def\bea{\begin{eqnarray}}
\def\eea{\end{eqnarray}}
\def\si{\sigma}
\def\br{{\bf r}}

\documentstyle[prb,aps,eqsecnum,multicol,epsf]{revtex}

\begin{document}
\draft
\title{Resonant Optical Nonlinearity of Conjugated Polymers}
\author{Ming-Che~Chang\thanks{changmc@phys.nthu.edu.tw} and
Hsin-Fei~Meng\thanks{meng@cc.nctu.edu.tw}}
\address{$^*$Department of Physics, National Tsing Hua University,
Hsinchu 300, Taiwan}
\address{$^\dagger$Institute of Physics, National Chiao Tung University,
Hsinchu 300, Taiwan}
\date{\today}
\maketitle

\begin{abstract}
{When the energy of a pump wave is in resonance with the exciton 
creation energy, the electric susceptibility of a conjugated polymer 
in response to the probe wave is altered by the 
exciton gas. In this paper, we calculate the dependence of this change 
on the the exciton populations by the equation of motion (EOM) method.  
The magnitude of optical nonlinearity is also influenced by 
ambient temperature, by the extent of exciton wave functions, and 
by the strength of electron-electron interaction. All of these 
factors can be easily incorporated in the EOM approach systematically. 
Using the material parameters for polydiacetylene (PDA), the optical Kerr 
coefficient $n_2$ obtained is about $10^{-8}\ {\rm cm}^2/{\rm W}$, 
which is close to experimental value, and is four orders of 
magnitude larger than the value in nonresonant pump experiments.}
\end{abstract} 

\begin{multicols}{2}
\section{Introduction}

As a class of materials promising for applications in all-optical
devices for communication and data processing,
conjugated polymers have been a subject of
great research interest.\cite{Phys} Conjugated
polymer such as PDA has long been recognized to exhibit 
large nonresonant third-order optical nonlinearities.\cite{Sauteret}
Most of the theoretical works on the nonlinear optics of these materials 
are based on the perturbative sum-over-state formula\cite{Garito} 
that suits {\it nonresonant} pumping well (but is inappropriate for resonant 
pumping). A fair agreement between experiments and theories has been achieved
for a variety of nonresonant spectra, including two-photon absorption,
third-harmonic generation, and electroabsorption.\cite{Abe}
In addition, EOM approach has also been employed in these 
studies.\cite{Mukamel_0} The EOM for polarization, which is similar to
the semiconductor Bloch equation originally used for inorganic
semiconductors\cite{Lindberg}, is a general tool whose validity goes 
beyond the perturbative regime\cite{Green1} and will be used in this paper.

For conjugated polymers under {\it resonant} pumping, most of the optically 
excited electrons and holes remain bound at room temperature and form an 
exciton gas, instead of a plasma as in the III-V 
semiconductors.\cite{Hartmut} This is because the magnitude of the 
exciton binding energy in conjugated polymers (of the order of 1 eV) is 
much larger than that in inorganic semiconductors (a few meV).
This presents a difficult situation for the calculation of resonant 
nonlinearity. Even though there are a lot of experimental 
works, few theory has been devoted to this subject.
A simple and heuristic explanation of 
this nonlinearity based on phase-space-filling (PSF) has been proposed, 
and the result agrees quantitatively with experiments.\cite{Green2} 
The esssence of PSF is that when there is a finite concentration of 
excitons, the phase space for further excitons to form is 
reduced because of the exclusion principle, and henceforth 
the probe absorption signal is reduced. The limitation of PSF is that it 
cannot predict the whole spectral response of the electric 
susceptibility, nor the effects of  
temperature, strength of electron-electron interaction ... etc. 

Motivated by the desire to have a more accurate theoretical tool to 
understand this phenomenon, we present a detailed analysis of the 
nonlinear susceptibility based on EOM. We will utilize the 
technique developed by Haug, Koch, and Schmitt-Rink to make a 
connection between exciton populations and the electron populations 
that appear in the EOM.\cite{Haug2} However, since the link is 
suitable for dilute exciton gas only, we shall concentrate our study in 
this regime. Furthermore, from the comparison between the 
lineshapes of time-resolved and cw photoluminescence spectra, 
the typical intraband scattering time is estimated to be less than
picosecond in polymers.\cite{Yan} Therefore, it is possible to 
consider a simpler situation where the
exciton gas is in quasi-thermal equilibrium, whose population is
determined by the intensity of the pump wave. Because of these 
approximations, phenomena such as exciton-exciton interaction,  
excited-state absorption, and off-equilibrium momentum distribution are 
not considered.

In this paper we have done extensive studies 
on the optical nonlinearity of PDA. There are several advantages in choosing 
this material: First, it 
has the largest nonresonant third-order nonlinearity of all polymers and 
very large resonant nonlinearity. Second, it is one of a few conjugated 
polymers that can form high quality single crystals and is more amenable to
theoretical analysis. Third, its chain-to-chain distance is large because of 
large side groups and therefore the inter-chain interaction is less
important. Various aspects of the nonlinear spectra are investigated. 
By choosing a reasonable value of the Coulomb 
interaction strength, the resonant optical Kerr coefficient $n_2$ being 
calculated agrees very well with experiments. Furthermore, the effect of 
electron-electron interaction on the height and position of the 
exciton absorption peak is studied. We also study the 
influence of temperature, as well as the relative populations of singlet and 
triplet excitons, on the probe spectra.
This work provides a systematic analysis of the 
influence of various microscopic parameters on the optical nonlinearity 
and may serve as a guide for the search of optical materials with 
larger resonant nonlinearities.

This paper is organized as follows. In section II, the equation of motion 
for polarization in conjugated polymers is derived. In section III, the 
Coulomb potential matrix elements and electron occupation numbers
are calculated. The numerical results are presented in 
section VI, and Sec.~V is the conclusion.

\section{Equation of Motion for Polarization}

A conjugated polymer is a macro-molecule with many electronic 
and ionic degrees of freedom. There have been several attempts to include
both types of degrees of freedom in 
calculations on the electronic and optical properties of 
conjugated polymers. This is a very difficult task. To date, an 
exact solution for chains longer than 20 sites is still 
lacking.\cite{Wen} It is especially challenging to calculate {\it resonant} 
optical nonlinearity by including the effects of electrons (including 
excitons) and phonons simultaneously. In this paper, we choose a modest 
approach and focus only on the electronic contribution to the nonlinearity.
By doing so, we will not, for example,
be able to produce satellite phonon peaks next to 
the main resonant absorption peak.\cite{Weiser} Our main 
goal is to calculate the optical nonlinearity due to the exciton gas, which is 
derived from the dependence of the magnitude of the main 
absorption peak on the exciton populations. 
In fact, Greene {\it et al} \cite{Green1,Green2} have demonstrated 
that the PSF model, whose origin is purely electronic, can explain major 
features of the spectra well. Furthermore, we consider only the 
dynamics of $\pi$-electrons; the $\si$-electrons are tightly bound to 
the ions and has little influence on the dynamical response. 
However, they do contribute to the renormalization of the interaction 
between $\pi$-electrons, and between ions and $\pi$-electrons. These 
effects appear implicitly through the parameters in the $\pi$-electron 
Hamiltonian.

Our calculation is based on the SSH-like Hamiltonian with 
electron-electron interactions, $H=H_0+H_1+H_2$, where \be
H_0=\sum_{k\si}\left(\epsilon_{ck}a_{ck\si}^\dagger a_{ck\si}
+\epsilon_{vk}a_{vk\si}^\dagger a_{vk\si}\right ),
\ee
\be
H_1=\sum_{\lambda_1\lambda_2\lambda_3\lambda_4}\sum_{k_1k_2k_3k_4}
{\cal V}_{\lambda_1\lambda_2\lambda_3\lambda_4}^{k_1k_2k_3k_4}
a_{\lambda_1 k_1\si}^\dagger a_{\lambda_2 k_2\si'}^\dagger
a_{\lambda_3 k_3\si'} a_{\lambda_4 k_4\si},
\ee
and 
\be
H_2=-E(t)\sum_{k\si}\left(\mu_{vc}(k)a_{vk\si}^\dagger a_{ck\si}
+\mu_{cv}(k)a_{ck\si}^\dagger a_{vk\si}\right ).
\ee
In these equations, $\epsilon_{c,vk}$ are the energies for the 
dimerization-induced conduction and valence band, 
${\cal V}_{\lambda_1\lambda_2\lambda_3\lambda_4}^{k_1k_2k_3k_4}$ are the 
Coulomb potential matrix elements ($\lambda=c,v$), and
$\mu_{\lambda\lambda'}(k)$ are the dipole matrix elements.
$H_2$ describes the coupling between polarization and the {\it probe} field 
$E(t)$ along the chain. The influence of the pump field will be
accounted for when calculating the conduction electron populations in the 
next section. 
The potential matrix elements are 
\end{multicols}
\be
{\cal V}_{\lambda_1\lambda_2\lambda_3\lambda_4}^{k_1 k_2 k_3 k_4}
= \int d^3\br 
d^3 \br'\Psi_{\lambda_1 k_1}^*(\br)\Psi_{\lambda_2 k_2}^*(\br')
V_{ee}(\br-\br')\Psi_{\lambda_3 k_3}(\br')\Psi_{\lambda_4 k_4}(\br),
\ee
\begin{multicols}{2}
\noindent where $V_{ee}$ is the screened Coulomb interaction between  the 
$\pi$-electrons, and $\Psi_{\lambda k}$ are the Bloch states solved from 
$H_0$.
The usual practice is to neglect, from the very beginning, the terms that do 
not conserve electron numbers in each band, which include half of 
the sixteen terms that do not have equal numbers of $c$ indices and $v$ 
indices, plus ${\cal V}_{vvcc}^{k_1 k_2 k_3 k_4}$ and ${\cal 
V}_{cvcv}^{k_1 k_2 k_3 k_4}$.\cite{Haug}
This is reasonable in metals or inorganic
semiconductors where the Coulomb interaction is weak, but is not 
necessarily valid in conjugated
polymers where the Coulomb interaction is much stronger, as indicated by 
the large exciton binding energies. Therefore, these 
terms will be kept in the derivation until they are proven negligible. 
It will be shown later that the terms without equal numbers of $c$ and 
$v$ indices indeed make no contribution, but the ${\cal 
V}_{vvcc}^{k_1k_2k_3k_4}$ and
${\cal V}_{cvcv}^{k_1k_2k_3k_4}$ terms cannot be ignored.  
In fact, unlike all the other terms,
these two terms do not conserve the electron
and hole spin individually, thus are essential to the lifting of the 
four-fold degeneracy in the spin subspace of the exciton states.

In the following, we derive the EOM for 
$p_{k\si}=\langle a^\dagger_{ck\si} a_{vk\si}\rangle$.
The total polarization $\langle P\rangle$, which is equal to $\sum_{k \si} 
(\mu_{vc}(k) p_{k \si} + \mu_{cv}(k) p^*_{k \si})$, can be easily 
obtained by integration over all $k$'s. In general, terms of the form
$\sum_{k_1k_2k_3}\sum_{\si'}{\cal V}_{\lambda_1\lambda_2\lambda_3\lambda}^
{k_1k_2k_3k} a_{\lambda_1k_1\si}^\dagger
a_{\lambda_2 k_2\si'}^\dagger a_{\lambda_3 k_3\si'} a_{\lambda k\si}$
are encountered in the derivation. By using the RPA,
the mean values of the product of four operators can be factorized,
\end{multicols}
\bea
\langle a^{\dagger}_{\lambda_1 k_1 \si} a^{\dagger}_{\lambda_2 k_2 \si'}
a_{\lambda_3 k_3 \si'} a_{\lambda_4 k_4 \si} \rangle &
= &
\langle a^{\dagger}_{\lambda_1 k_1 \si} a_{\lambda_4 k_1 \si} \rangle
\langle a^{\dagger}_{\lambda_2 k_2 \si'} a_{\lambda_3 k_2 \si'} \rangle
\delta_{k_1 k_4} \delta_{k_2 k_3}  \cr
& - &
\langle a^{\dagger}_{\lambda_1 k_1 \si} a_{\lambda_3 k_1 \si} \rangle
\langle a^{\dagger}_{\lambda_2 k_2 \si} a_{\lambda_4 k_2 \si} \rangle
\delta_{k_1 k_3} \delta_{k_2 k_4} \delta_{\si \si'}.
\eea
After a straightforward but tedious calculation, we have
\bea\label{main0}
i\hbar\frac{\partial p_{k\si}}{\partial t}
&=& (\epsilon_{vk}-\epsilon_{ck})p_{k\si}
+E(t) \mu_{vc}(k)(n_{vk\si}-n_{ck\si})\cr 
&-& \sum_{k'} 
\left[ \left({\cal V}_{vvvv}^{kk'}-{\cal V}_{cvcv}^{kk'}\right)
(n_{vk'\si}-n_{ck'\si})
-\sum_{\si'}\left(\tilde{\cal V}_{vvvv}^{kk'}
-\tilde{\cal V}_{cvvc}^{kk'}\right)\left(n_{vk'\si'}-n_{ck'\si'}\right)
\right ]p_{k\si}\cr
&+& \sum_{k'}\left[ 
{\cal V}_{vccv}^{kk'}p_{k'\si}
+{\cal V}_{vvcc}^{kk'}p_{k'\si}^*
-\sum_{\si'}\left(
\tilde{\cal V}_{vcvc}^{kk'}p_{k'\si'}
+\tilde{\cal V}_{vvcc}^{kk'}p_{k'\si'}^*\right )
\right ](n_{vk\si}-n_{ck\si})\cr
&+& \sum_{k'}\left[ 
{\cal V}_{vvcv}^{kk'}n_{vk'\si}
+{\cal V}_{vccc}^{kk'}n_{ck'\si} 
-\sum_{\si'}\left(
\tilde{\cal V}_{vvvc}^{kk'}n_{vk'\si'}
+\tilde{\cal V}_{vccc}^{kk'}n_{ck'\si'}\right )
\right ](n_{vk\si}-n_{ck\si}),
\eea
\begin{multicols}{2}
\noindent where $n_{c,vk\si} = 
\langle a^{\dagger}_{c,v k \si} a_{c,v k \si} \rangle$ 
are the one electron occupation numbers in conduction 
or valence band, ${\cal V}_{\lambda_1 \lambda_2\lambda_3 
\lambda}^{kk'}$ and $\tilde{\cal V}_{\lambda_1 \lambda_2\lambda_3
\lambda}^{kk'}$ are abbreviations for the Coulomb terms, ${\cal 
V}_{\lambda_1 \lambda_2\lambda_3 \lambda}^{kk'kk'}$, and 
the exchange terms, ${\cal V}_{\lambda_1 \lambda_2\lambda_3 
\lambda}^{kk'k'k}$, respectively.
The terms that are quadratic in $p_{k\si}$ are neglected in 
Eq.~(\ref{main0}), because only the linear response of the probe wave is 
considered. Also, relations such as ${\cal V}_{cccc}^{kk'}
={\cal V}_{vvvv}^{kk'}$ and ${\cal V}_{cvcv}^{kk'}={\cal V}_{vcvc}^{kk'}$
have been used, which are based on the symmetry between conduction band 
and valence band in the present model.
Note that the potential matrix 
elements with unequal numbers of $c$ and $v$ indices result 
in the last line of Eq.~(\ref{main0}). These matrix elements are 
multiplied by terms quadratic in electron or hole occupations, which 
under resonant pumping may be large. 
Notwithstanding, because of inversion symmetry, it can be shown 
that for thermalized exciton gas, these terms actually have no effect on 
the dynamics (see Appendix A). Consequently, Eq.~(\ref{main0}) becomes 
\end{multicols}
\bea\label{main1}
& &\left[(\epsilon_{ck}-\epsilon_{vk})
-\omega-i\gamma \right]\bar p_{k\sigma}(\omega)\cr 
&=& E(\omega)\mu_{vc}(n_{vk\sigma}-n_{ck\sigma})\cr
&-& \sum_{k'}\left[ 
\left({\cal V}_{vvvv}^{kk'}-{\cal V}_{cvcv}^{kk'}\right)
\left(n_{vk'\sigma}-n_{ck'\sigma}\right)
-\sum_{\si'}\left(\tilde{\cal V}_{vvvv}^{kk'}
-\tilde{\cal V}_{cvvc}^{kk'}\right)\left(n_{vk'\si'}-n_{ck'\si'}\right)
\right]\bar p_{k\sigma}(\omega)\cr
&+& \sum_{k'}\left[ 
{\cal V}_{vccv}^{kk'} \bar p_{k'\sigma}(\omega)
+{\cal V}_{vvcc}^{kk'} \bar p_{k'\sigma}^*(-\omega)
-\sum_{\si'}\left(
\tilde{\cal V}_{vcvc}^{kk'}\bar p_{k'\si'}(\omega)
+\tilde{\cal V}_{vvcc}^{kk'}\bar p_{k'\si'}^*(-\omega)\right )
\right]
(n_{vk\sigma}-n_{ck\sigma}), 
\eea
\begin{multicols}{2}
\noindent where $\bar p_{k\sigma}= (p_{k\sigma}+p_{-k\sigma})/2$, and a damping 
term $i\gamma$ has been added. Notice that $\bar p_{k\si}^*(-\omega)\ne
\bar p_{k\si}(\omega)$ because $\hat p_{k\si}$ is not Hermitian. 
Therefore, Eq.~(\ref{main1}) has to be solved in 
conjunction with the equation satisfied by $\bar p^*_{k\si}$, which is 
similar to Eq.~(\ref{main1}) but with $\omega$ replaced by $-\omega$ and 
$\bar p_{k\si}$ replaced by $\bar p^*_{k\si}$.
The meaning of the various parts on the right hand side of the equation is 
explained below:
The first square bracket, after being summed over $k'$,
contributes to the band gap renormalization. The magnitude of 
renormalization 
depends on the strength of inter-electron interaction, {\it as well as} the 
electron populations. Inside the second square bracket, the ${\cal 
V}_{vccv}^{kk'}$ term is most crucial to the formation of excitons;
the ${\cal V}_{vvcc}^{kk'}$ term is related to the singlet-triplet splitting 
of the exciton levels and leads to an unusual coupling between positive 
and negative frequency components of the polarization. For the exchange 
terms, it will be shown in the next section that $\tilde{\cal V}_{vvvv}^{kk'}
=\tilde{\cal V}_{cvvc}^{kk'}$ because of charge neutrality
condition. Therefore, there is no exchange effect in the first square 
bracket. We can also show that  $\tilde{\cal 
V}_{vcvc}^{kk'}$ and $\tilde{\cal V}_{vvcc}^{kk'}$ are simply constants 
and can be treated as corrections to ${\cal 
V}^{kk'}_{vccv}$ and ${\cal V}^{kk'}_{vvcc}$. The only unknowns in Eq.~(\ref{main1}) are $\{\bar 
p_{k\si}(\omega), \bar p_{k\si}^*(-\omega)\}$. All of the other 
quantities, including the potential matrix elements and the electron 
populations, can be obtained given physical conditions such as the strength 
of electron interaction, the intensity of pump wave\dots etc. This is 
derived in the next section.

\section{Potential matrix element and electron population}

\subsection{Potential matrix element}

A natural choice for the interaction potential between electrons
is $V_0/|\br-\br'|$, where $V_0$ is given by 
$e^2/\epsilon a_0$, $\epsilon$ is the intra-chain dielectric constant, 
and $a_0$ is the average distance between neighboring sites.
The position $r$, being dimensionless now, is measured in units of $a_0$.
To calculate ${\cal V}_{\lambda_1\lambda_2 \lambda_3\lambda_4}^{kk'}$,
the unperturbed eigenstates are expanded by localized Wannier 
functions,\cite{Abe2} 
\be
\Psi_{\lambda k}(\br)=\sum_{j=1,2} u_{\lambda j}(k)\left(\frac{1}{\sqrt M}
\sum_{m=1}^M e^{2ikm}W_{2(m-1)+j}(\br) \right),
\ee
where $M$ is the number of unit cells. 
The total chain length is $2M$.
Defining $\zeta_k=\left[e^{ik}(t_0\cos
k-i\delta t\sin k)/\epsilon_{ck} \right ]^{1/2}$, where $t_0-(-1)^l 
\delta t$ is 
the hopping amplitude between neighboring sites, then 
\be
\left(
\begin{array}{ll}
u_{c1}(k) & u_{c2}(k) \\
u_{v1}(k) & u_{v2}(k)
\end{array}
\right)
=
\left(
\begin{array}{rr}
\zeta^*_k & \zeta_k \\
-\zeta^*_k & \zeta_k
\end{array}
\right).
\ee
When calculating the matrix elements of $V_{ee}(\br-\br')$, only the
integrals involving Wannier functions at the same site are kept
(zero differential overlap approximation).\cite{Hartmann}
To improve upon this, we need to know the shape of atomic orbitals, which
will not be considered here. For the same site, 
there is a finite on-site energy $U_0=1/2\int d^3\br 
d^3\br' W^*(\br)W^*(\br')V_{ee}(\br - \br')W(\br')W(\br)$. 
Since the exact form of the Wannier function is not known, $U_0$ is 
treated as a parameter independent of $V_0$. So our choice of the 
potential is essentially of the Pariser-Parr-Pople form. Defining 
\bea
V_1(q)&=&\frac{1}{M}\sum_m V_{ee}(2m+1)e^{-i(2m+1)q},\cr
V_2(q)&=&\frac{1}{M}\sum_m V_{ee}(2m)e^{-2imq},
\eea
then a straightforward calculation gives
\end{multicols}
\bea\label{V12}
{\cal V}_{\lambda_1\lambda_2 \lambda_3\lambda_4}^{kk'}
&=& V_2(q)\left[ \sum_{j=1}^2 u_{\lambda_1 j}^*(k) u_{\lambda_2 j}^*(k')
u_{\lambda_3 j}(k) u_{\lambda_4 j}(k') \right]\cr
&+& V_1(q)\left[ e^{iq} u_{\lambda_1 2}^*(k) u_{\lambda_2 1}^*(k')
u_{\lambda_3 1}(k) u_{\lambda_4 2}(k')
+e^{-iq} u_{\lambda_1 1}^*(k) u_{\lambda_2 2}^*(k')
u_{\lambda_3 2}(k) u_{\lambda_4 1}(k') \right],\cr
\tilde{\cal V}_{\lambda_1\lambda_2 \lambda_3\lambda_4}^{kk'}
&=& \frac{1}{2}V_2(0)\cr
&+& V_1(0)\left[ u_{\lambda_1 2}^*(k) u_{\lambda_2 1}^*(k')
u_{\lambda_3 1}(k') u_{\lambda_4 2}(k)
+u_{\lambda_1 1}^*(k) u_{\lambda_2 2}^*(k')
u_{\lambda_3 2}(k') u_{\lambda_4 1}(k) \right],
\eea
\begin{multicols}{2}
\noindent where $q= k-k'$. We require $V_1(0)+V_2(0)=0$ because of the 
charge 
neutrality condition. After combining Eq.~(3.4) with Eq.~(3.2), it can be 
shown that $\tilde{\cal V}_{vvvv}^{kk'}-\tilde{\cal 
V}_{cvvc}^{kk'}=(V_2(0)+V_1(0))/2=0$, and $\tilde{\cal V}_{vcvc}^{kk'}
=\tilde{\cal V}_{vvcc}^{kk'}=(V_2(0)-V_1(0))/2$. 
We emphasize that this nonzero exchange correction is crucial to the 
reduction of the exciton binding energy when $U_0$ is tuned to a larger 
value while keeping $V_0$ fixed. The same behavior is observed in Abe's 
paper\cite{Abe2}, where their concern is the energy spectrum of the 
excitons.\cite{U_0} 

\subsection{Electron population}

The electron population can be linked to the exciton population by using 
the method developed by Haug, Koch, and Schmitt-Rink. Their relation is 
derived below. Firstly, the connection between electron operators and an 
exciton creation operator $e^\dagger_{njmK}$ is\cite{Haug2}
\be
a_{ck_1\si_1}^\dagger a_{vk_2\si_2}=\frac{1}{\sqrt{2M}}\sum_{njm}
\langle jm|\si_1\si_2\rangle \phi_{njm}^*(k) e^\dagger_{njmK} 
\, ,
\ee
where $\langle jm|\si_1\si_2\rangle $ is the Clebsh-Gordon coefficient,
$k=(k_1+k_2)/2$, $K= k_1-k_2$, and
$\phi_{njm}(k)$ is the wave function of an exciton at the $n$-th
bound state with angular momentum $\{jm\}$.
Helped by this relation, we can write $n_{c,vk}$ in terms of 
exciton operators as follows:
By using an identity operator 
\be
I=\frac{1}{N_c} ( N - \sum_{k \si} a_{vk\si}^\dagger a_{vk\si} ) =
\frac{1}{N_c}\sum_{k\si}a_{vk\si} a_{vk\si}^\dagger ,
\ee
where $N$ is the total number of electrons and $N_c$ is the number of 
conduction electrons, we have 
\bea
\hat n_{ck_1}&\equiv & 
\sum_{\si_1} \hat n_{ck_1\si_1}\cr
&=& \frac{1}{N_c} \sum_{k_2\si_1\si_2} 
a_{ck_1\si_1}^\dagger a_{ck_1\si_1}a_{vk_2\si_2} a_{vk_2\si_2}^\dagger\cr
&=& \frac{1}{2MN_c}\sum_{nn'jm}\sum_{k_2} \phi_{njm}^*(k)\phi_{n'jm}(k)
e_{njmK}^\dagger e_{n'jmK}.
\eea

For a dilute exciton gas, the exciton number $\langle e_{njmK}^\dagger 
e_{n'jmK} \rangle \simeq g_{njmK}\delta_{nn'}$, where $g_{njmK}$ is the 
thermal distribution of excitons.\cite{Haug2} Therefore,
\be
n_{ck_1}= \frac{1}{2MN_c}\sum_{nk_2}
\left(|\phi_n^s(k)|^2 g_{nK}^s+3 |\phi_n^t(k)|^2 g_{nK}^t \right ), 
\ee
where the subscripts $s$ and $t$ stand for `singlet' ($j=0,m=0$) and 
`triplet' ($j=1, m=\pm 1,0$ ) respectively; the factor 3 in front of 
$g_{nK}^t$ accounts for the triplet degeneracy. 
For a one-dimensional system, the exciton wave function near band buttom  
can be approximated by\cite{Green1} 
$\phi_0^{s,t}(k)=( 2r_0^{s,t}/\pi )^{1/2}/
\{1+\left[(k\pm\pi/2)r_0^{s,t}\right]^2\}$,
where $r_0^{s,t}$ are the radii of the excitons. Note that
the conduction band bottoms are at $\mp\pi/2$. The ground state makes 
major contribution since all the other levels are far off resonance and 
barely occupied. According to Abe's single configuration interaction (SCI) 
calculation,\cite{Abe2} the ratio $U_0/V_0$ determines whether the 
singlet exciton level (${\ }^1B_u$) or the triplet exciton level 
(${\ }^3B_u$) is lower in energy. They found that, for $V_0=t_0$ and 
$U_0>1.39 V_0$, ${\ }^3B_u$ is lower than ${\ }^1B_u$, 
and vice versa. All the even-parity $A_g$ states are lying at higher 
energies. However, it is found in 
some finite-chain calculations that the lowest excitation is actually an 
even parity state.\cite{even} This would have a significant effect on the 
efficiency of luminescence since the optically excited electrons 
at ${\ }^1B_u$ may relax to the  $A_g$ state first, then release 
their energy via non-radiative channels. Nontheless, it has been shown 
that the relaxation rate from the 1$\ ^1B_u$ exciton to the $2\ ^1A_g$ 
exciton is much smaller than the relaxation rate to the ground 
state.\cite{Koby} Furthermore, from the point of view of phase space 
filling, both $1\ ^1B_u$ and $2\ ^1A_g$ excitons contribute almost 
equally to the reduction of phase space. Both factors seem to diminish 
the effect of this even-parity state. In fact, for the optical nonlinearity 
being studied, one study shows that, for both SCI and finite-chain 
calculations, the optical nonlinearity is determined almost entirely by the 
odd-parity ${\ }^1B_u$ exciton, a dominant $A_g$ exciton above the ${\ 
}^1B_u$ level (not considered here), and the threshold of the conduction 
band.\cite{Chandross} Therefore, as far as optical nonlinearity is concerned,
we will neglect the influence of this even-parity state in this paper.

The thermal distribution of excitons is given by  
$g_{nK}^{s,t}=(\exp\{\beta [\epsilon_n^{s,t}(K)-\mu_{s,t}]\}-1)^{-1}$,
where $\epsilon_n^{s,t}(K)=\epsilon_n^{s,t}(0)+\hbar^2K^2/2m_{ex}$ are the 
energies of excitons, the exciton effective mass $m_{ex}$ is 
$\hbar^2/a_0^2 t_0$, and $\mu_{s,t}$ are the quasi-chemical potentials. 
Immediately after the optical pumping, there 
are only singlet excitons because of the selection rule. Part of these
excitons then fall down to the triplet level via spin-orbital interaction.
Their populations are controlled by the quasi-chemical potentials in our 
calculation.

After summing over all the electrons in the conduction band, we have
\bea\label{hehe}
N_c&=& \sum_{k_1} \hat n_{ck_1}\cr 
&=& \frac{1}{2MN_c}\sum_{nn'jm}\sum_{kK} \phi_{njm}^*(k)\phi_{n'jm}(k)
e_{njmK}^\dagger e_{n'jmK}\cr
&=& \frac{1}{N_c} \sum_{njmK}e_{njmK}^\dagger e_{njmK},
\eea
where we have used the completeness relation for the exciton wave functions.
By taking the expectation value of Eq.~(\ref{hehe}), we have
\be
\sum_{njmK} \left\langle e_{njmK}^\dagger e_{njmK}\right\rangle
=N_c^2.
\ee
This identity is used to determine the values of the chemical potentials 
$\mu_s$ and $\mu_t$, once the total population of, and the relative 
populations between singlet and triplet excitons are given.

\section{Absorption Spectra: Numerical Result}

Most of this section is devoted to the calculation of  
the resonant nonlinear optical spectra of PDA for reasons 
stated at the introduction. At the end of this section we will comment 
briefly on the calculation for PPV. 

PDA has four carbon atoms per unit cell, and consequently four bands
in the tight-binding approximation.
Since our focus is on the exciton state within the 
band gap, which is composed of the electron and hole from the
middle two bands (conduction and valence band), the outer two bands
can be safely neglected. 
We choose the dimerization-induced band gap, 4$\delta t$, to be the 
unit of energy. The value of $t_0$, which determines the total band 
width of conduction band and valence band, is 
chosen to be 1.25.\cite{weak} The average bond length of PDA is 1.35 ${\rm 
\AA}$.\cite{Abe2}  The Coulomb interaction parameter $V_0=e^2/\epsilon 
a_0$ can be determined from the intra-chain dielectric constant 
$\epsilon$. According to the Kramers-Kronig analysis of the reflectivity 
and with permittivity measurements, this value is close to 3.\cite{Suhai} 
We choose $\epsilon =3.5$, such that $V_0=2.84$ eV, to fit the calculated 
exciton binding energy with the value observed in experiments. 
In most of the following calculations the on-site energy $U_0=2V_0$.\cite{Abe2}
We have also done several calculations using different choices of $U_0$ and 
$V_0$ values.(see Fig. 4) In all of the following calculations the 
number of sites is 400 and the damping $\gamma$ is 0.02. We have done 
calculations on a larger system with 800 sites and confirmed that the 
finite size effect is unimportant.

Figure 1 displays the electric susceptibility $\chi(\omega)$, in room 
temperature, for various concentrations of the singlet exciton gas.
The percentages of electrons excited from the valence band by the 
pump wave are indicated in the legend. The absorption peak in Im 
$\chi(\omega)$ can be clearly identified at $\hbar \omega=1.39$ in the 
absence of pumping. (the solid line) The renormalized band gap is 
determined by the first maximum of Im $\chi(\omega)$ beyond the exciton 
peak, which is barely observable at $\hbar \omega=1.71$. 
Notice that the band gap renormalization due to the
Coulomb interaction is quite large. With different exciton populations, the 
magnitude of renormalization also changes. (see the discussion after 
Eq.~(\ref{main1})) The exact value of the conduction band edge is 
difficult to measure experimentally because most of the oscillator 
strength is `concentrated' on the exciton peak. On the other hand, the 
singlet exciton absorption peak at $1.97$ eV is one of the few values 
that can be determined accurately and is generally agreed upon by 
researchers.\cite{Green2,Weiser,Koby} Therefore, it is used to set
the overall energy scale and that means $1.39\times 4\delta t=1.97$ eV, or 
$4\delta t=1.42$ eV. Consequently, the position of conduction band edge is at
2.43 eV in our calculation and the binding energy of the singlet exciton is 
approximately 0.46 eV. This falls within the range 0.4 -- 0.5 for the 
values reported. \cite{Green2,Weiser}  

\begin{figure}[hbt]
\centerline{\epsfysize=7cm \epsfxsize=6cm
\epsfbox{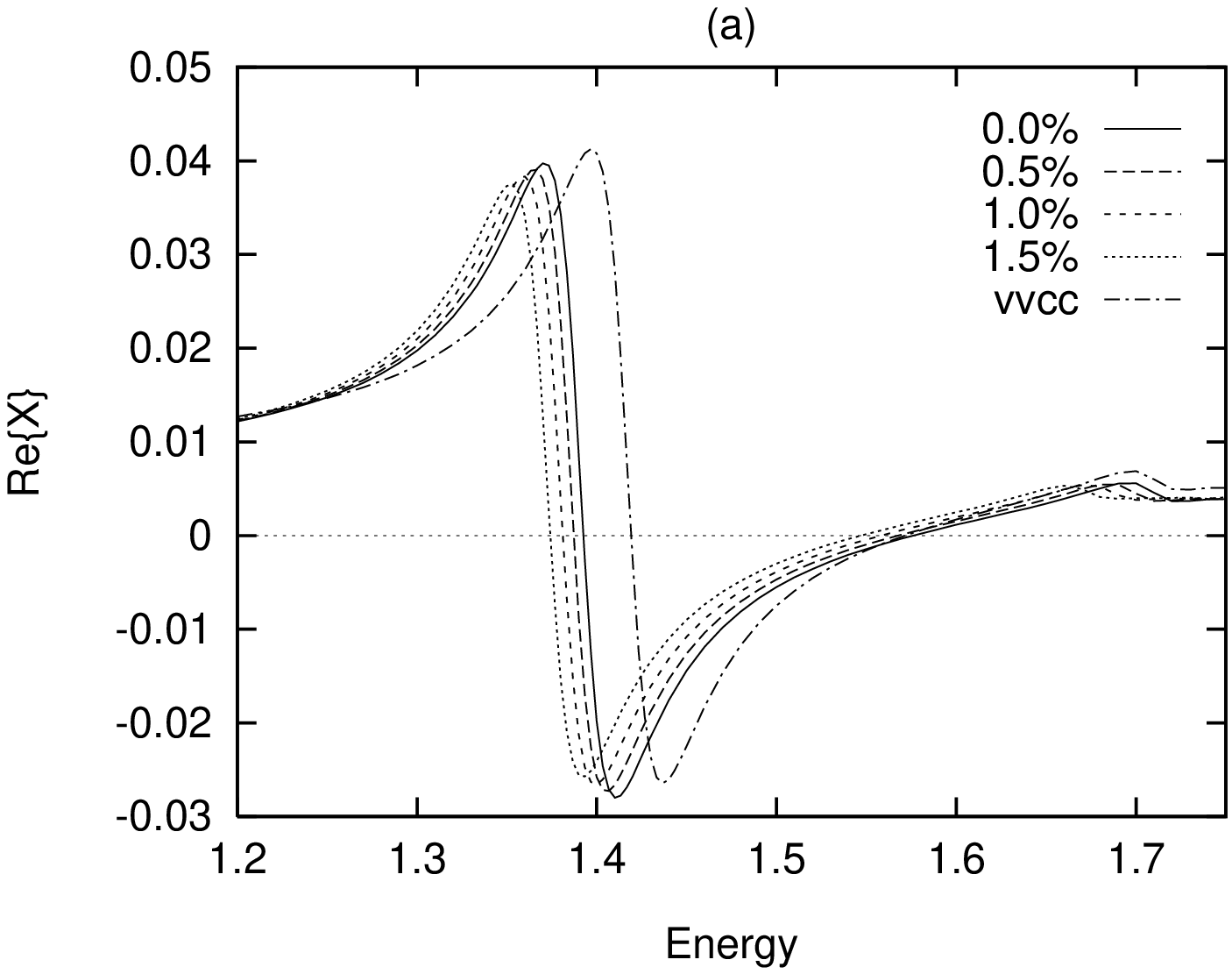}}
\centerline{\epsfysize=7cm \epsfxsize=6cm
\epsfbox{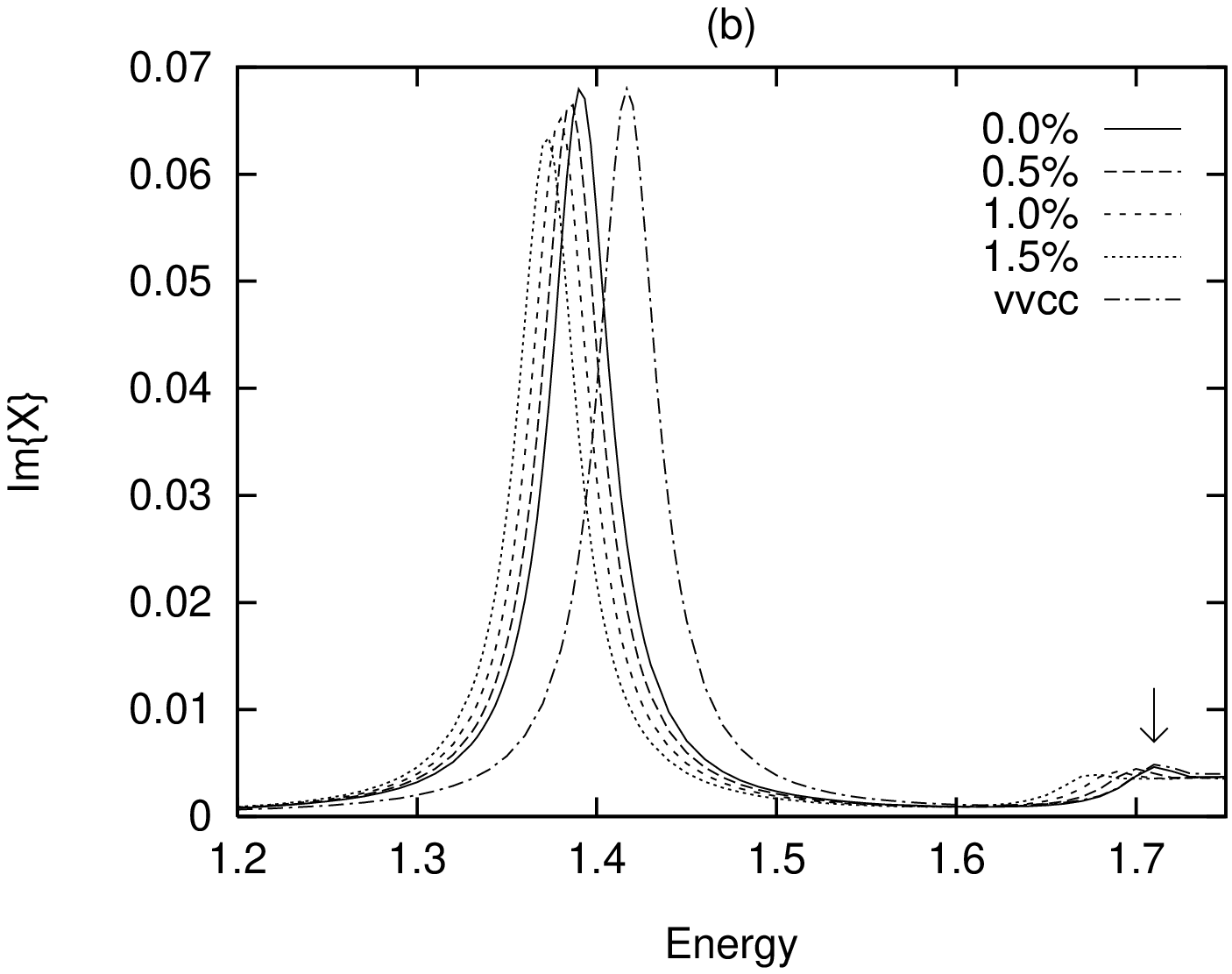}}
{Fig 1: Real (a) and imaginary (b) parts of
$\chi(\omega)$ for different exciton concentrations, labeled by the
percentages of the occupation of conduction band
(0, $0.5\%$, $1.0\%$, and $1.5\%$) at room temperature. Probe energy is
in units of the bare band gap, $4\delta t$.
The `vvcc' curve comes from the calculation without
the ${\cal V}^{kk'}_{vvcc}$ and $\tilde {\cal V}^{kk'}_{vvcc}$ terms and
in the absence of excitons. The arrow at the bottom right of
Fig.1(b) indicates the position of the band edge for the
solid line.}
\end{figure}

In Fig.~1, the absorption signal is reduced as exciton concentration 
increases. The reduction is approximately proportional to the number of 
excited electrons. This is consistent with the picture of PSF,
which gives \cite{Green2}
\be\label{PSF}
\delta f/f=- n_{ex}/n_{ex}^{sat} \, ,
\ee 
where $f$ is the oscillator strength, $n_{ex}$ is the exciton 
density per unit length, and $n_{ex}^{sat}$ is the saturation density, 
at which exciton wave functions begin to overlap in space.
By using Eq.~(\ref{PSF}), we can estimate the exciton radius and the 
result is about five unit cells, which is again consistent with 
earlier calculations.\cite{Weiser,Abe2} 

\begin{figure}[hbt]
\centerline{\epsfysize=7cm \epsfxsize=6cm
\epsfbox{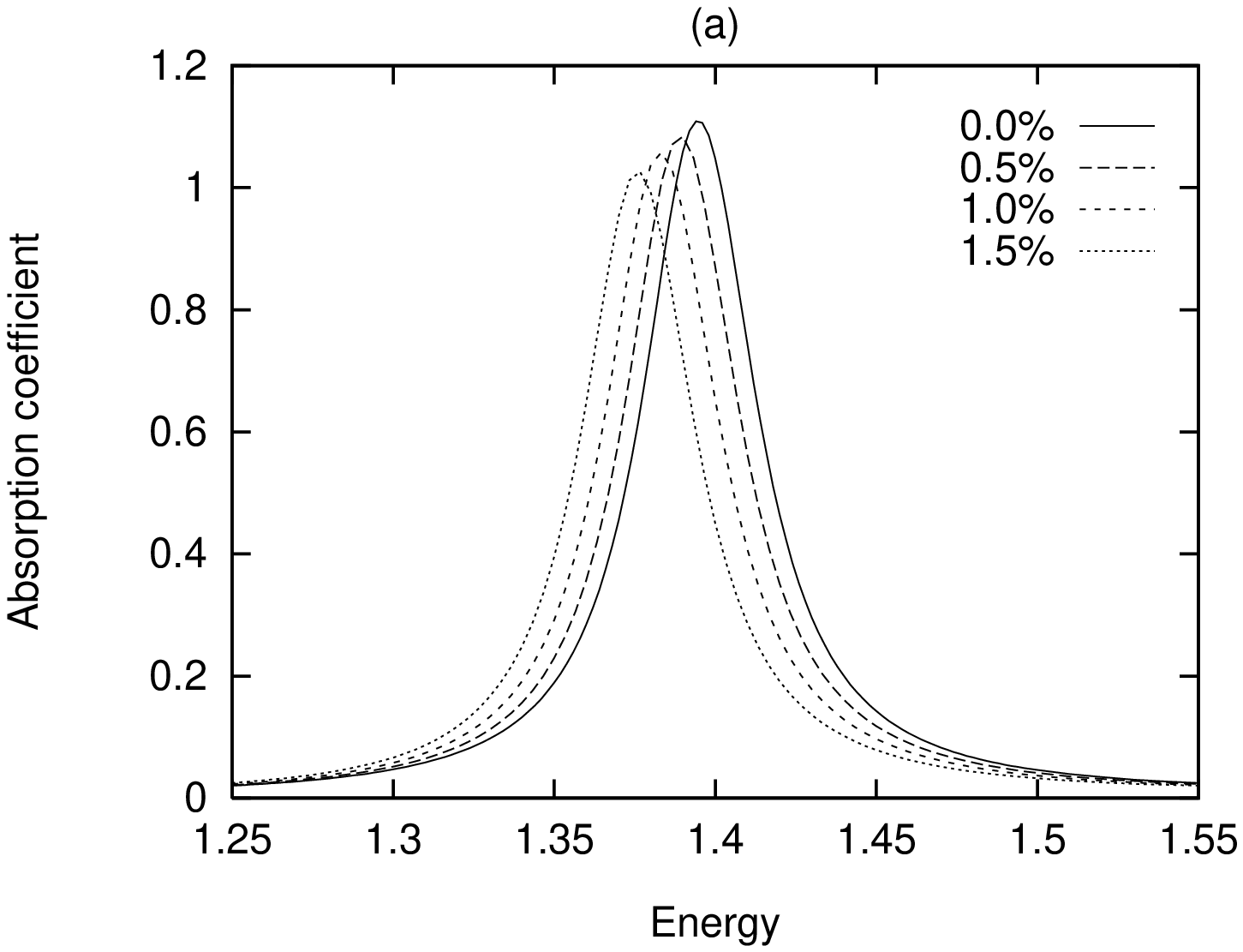}}
\centerline{\epsfysize=7cm \epsfxsize=6cm
\epsfbox{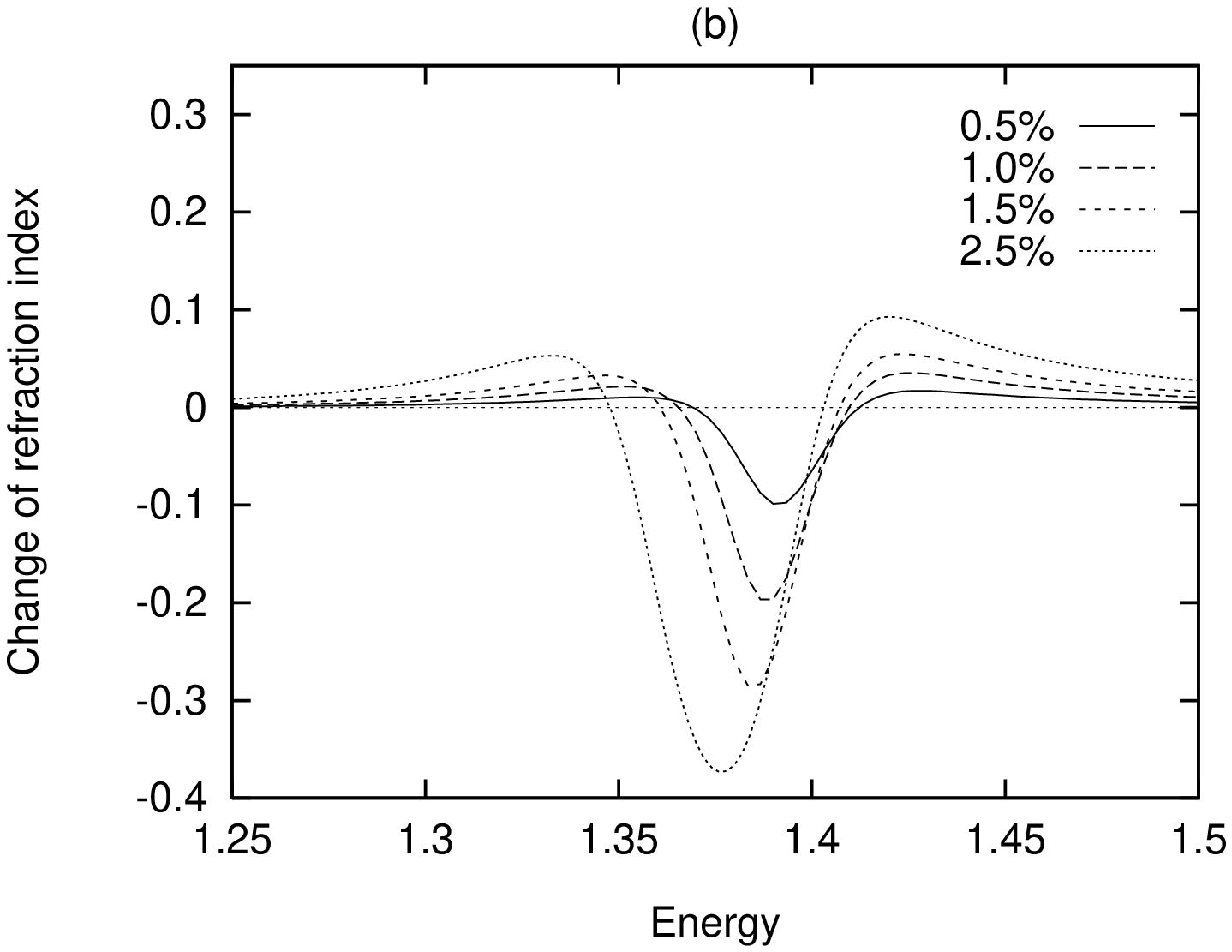}}
{Fig 2: The effect of different exciton concentrations on (a)
absorption coefficient $\alpha$ and (b) the change of
refraction index $\delta n$. $\alpha$ is in
units of $2.0\times 10^4\ {\rm cm}^{-1}$. The energy is in units of
$4\delta t$.
}
\end{figure}
 
Figure 2 shows the absorption coefficient $\alpha$ and the change of 
refraction index $\delta n$ near the exciton peak.
The values of the exciton concentration are the same as in Fig.~1. 
The chain-to-chain distance for PDA in the crystalline phase is about 
$10{\rm \AA}$,\cite{Green2} and the inter-site 
distance is $1.4{\rm\AA}$. Therefore, 1 \% concentration has $7.1\times 
10^{19}\ {\rm cm}^{-3}$ conduction 
electrons. The pump intensities $I_p$ required for this electron 
concentration $n_{ex}$ can be obtained from $I_p=n_{ex}\hbar 
\omega_p/\alpha_t \tau $ (reflection from the sample is ignored), where   
$\hbar\omega_p=1.97$ eV, the peak 
absorption $\alpha_t$ (including $\pi$-electrons and the background) is 
approximately $10^6\ {\rm cm}^{-1}$,\cite{Green1} and the recombination time 
$\tau$ is 2 ps.\cite{Green2} Consequently, to excite $7.1\times
10^{19}\ {\rm cm}^{-3}$ electrons requires a pump wave with 
intensity $I_p=1.14 \times 10^7\ {\rm W}/{\rm cm}^2$. 
The optical Kerr coefficient $n_2$, which measures the change of refraction 
index due to pumping, is given by $n_2 = |\delta n| / \delta I_p$.
For a pulse with $I_p=1.14 \times 10^7\ {\rm W}/{\rm cm}^2$, we have 
$|\delta n|=0.196$ at resonance (see Fig.2b)
and therefore $n_2=1.7 \times 10^{-8}\ {\rm cm}^2/{\rm W}$. 
This is four orders of magnitude larger than the nonresonant value,
and is close to Greene {\it et al}'s observation $3.0\times 
10^{-8}\ {\rm cm}^2/{\rm W}$.\cite{Green2}

\begin{figure}[hbt]
\centerline{\epsfysize=7cm \epsfxsize=6cm
\epsfbox{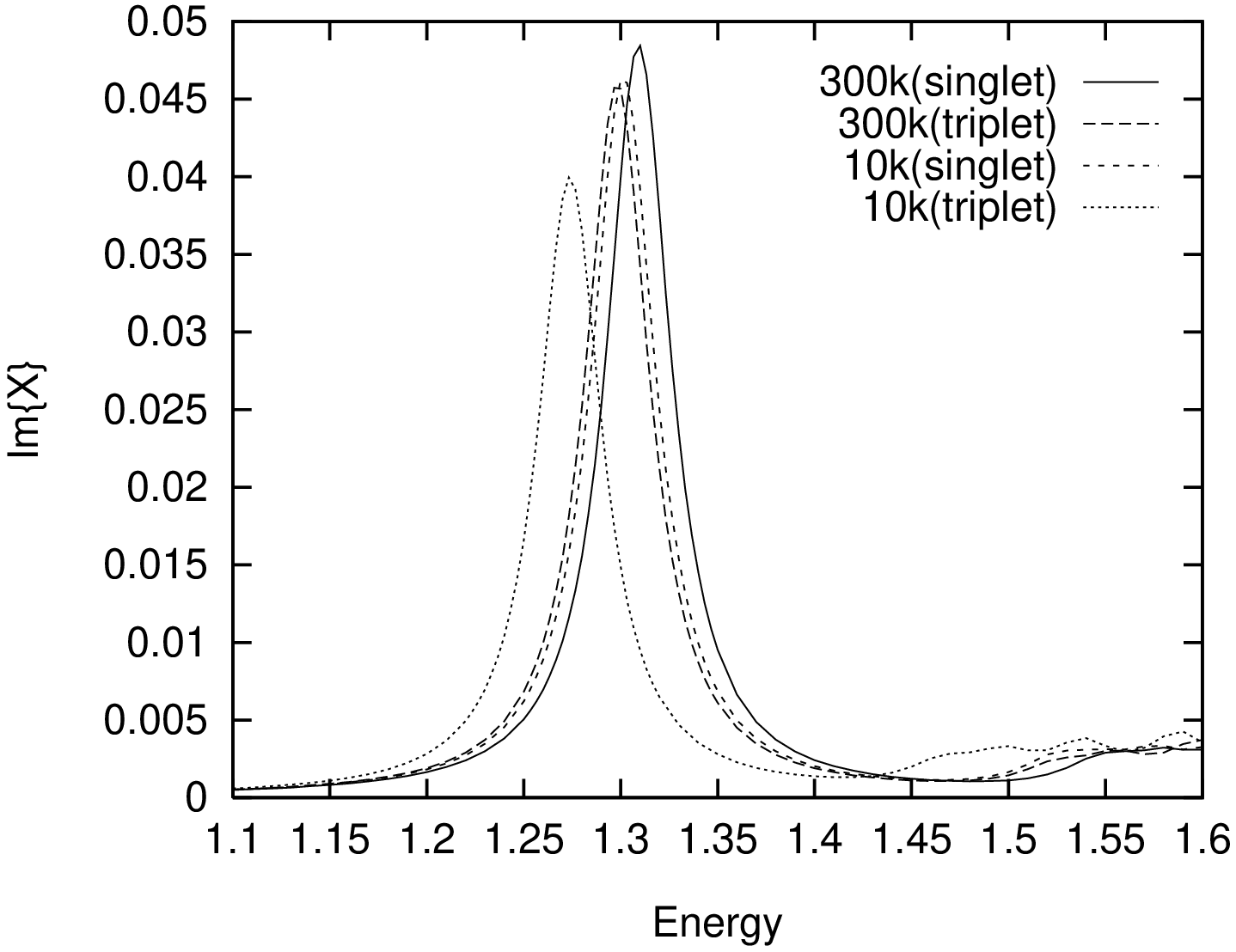}}
{Fig 3: The effect of temperatures and singlet/triplet
populations on Im $\chi(\omega)$. The percentage of conduction electrons
is $2.5\%$ for all of the curves. At room temperature, it
makes small difference whether the excitons are singlet or triplet.  The
distinction is more apparent at 10 K.
}
\end{figure}

All of the above calculations are for singlet excitons at room temperature. 
Using the same formalism it is quite easy to investigate the influence 
of exciton species and temperature on the resonant 
nonlinearity. We consider a two-level model where only the singlet (${\ 
}^1B_u$) and triplet (${\ }^3B_u$) excitons are considered. 
The exciton radii are chosen to be $6 a$ ($r_0^s$) and $4.5 a$
($r_0^t$).\cite{Abe2} In Figure 3, we show 
the extreme cases where the populations are either all-singlet 
or all-triplet. This difference traces back to 
the different distributions of the relative part of the electron-hole pair 
wave functions $\phi_0^{s,t}(k)$. It can be seen that the difference 
between singlet and triplet curves are more significant at low
temperature (10K). Such a temperature effect has not been studied 
experimentally, however. 

In Fig.~4, we show how different choices of the strength of the 
electron-electron interaction may affect the absorption spectra. It can 
be seen that the magnitude of 
$V_0$, the long-range interaction, has significant effect on the band gap
renormalization and the binding energy. 
For example, the band edges for $U_0=4$ and $V_0$= 1.5, 2, 2.5
are at 1.54, 1.71, and 1.89, respectively (see arrows in the figure).
The widths of band gap vary roughly linearly with $V_0$.
On the other hand, the short-ranged on-site energy $U_0$ has little 
influence on the bandgap. Furthermore, contrary to the effect of 
$V_0$, a larger $U_0$ leads to a smaller binding 
energy. This adverse effect can be traced back to the exchange terms 
$\tilde{\cal V}_{\lambda_1\lambda_2\lambda_3\lambda_4}^{kk'}$ in 
Eq.~(\ref{V12}). The dependence of the position of the exciton level on 
$U_0$ and $V_0$ resembles closely to the calculations by Abe {\it et 
al}.\cite{Abe2}    

\begin{figure}[hbt]
\centerline{\epsfysize=7cm \epsfxsize=6cm
\epsfbox{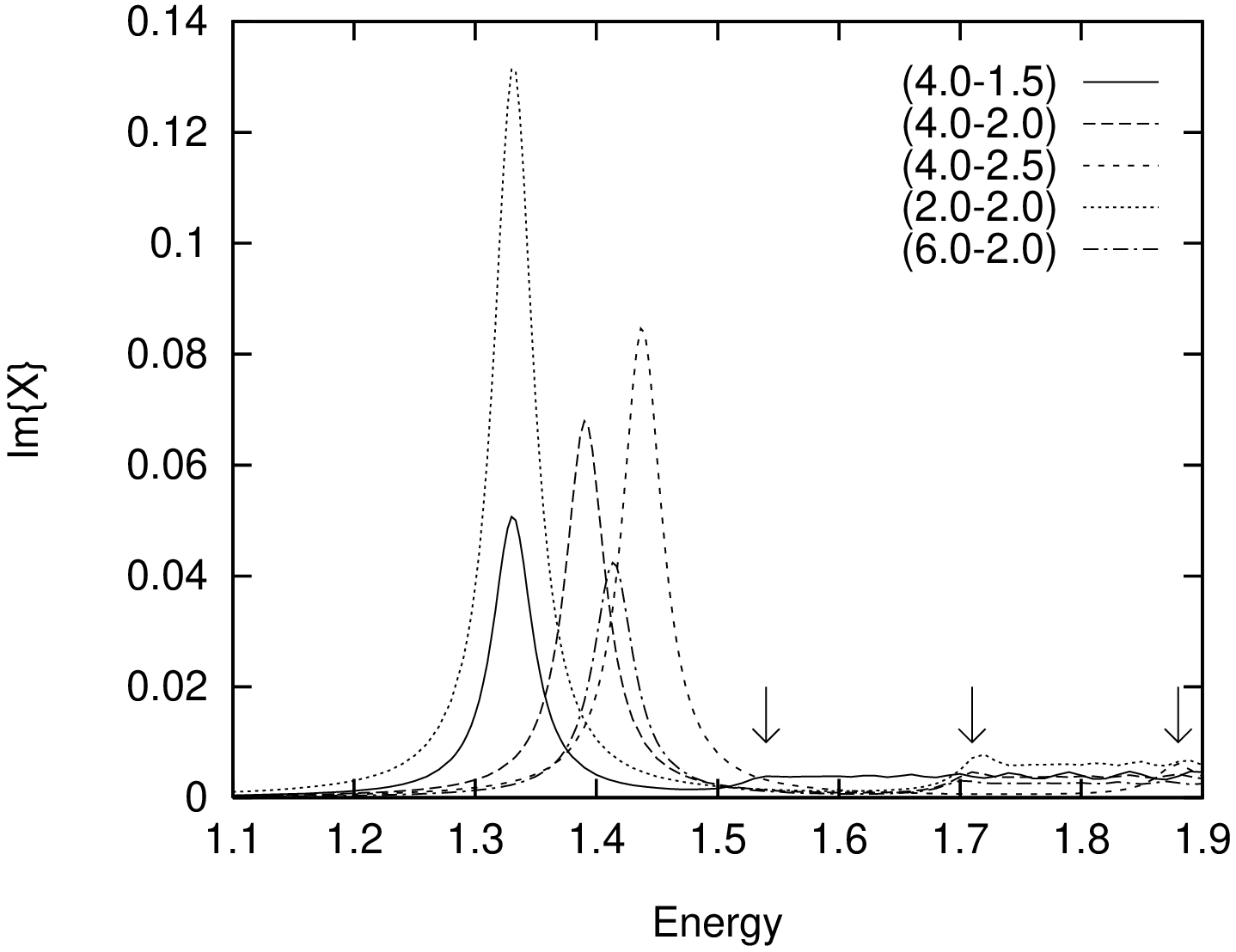}}
{Fig 4: The influence of Columbic parameters, $U_0$ and $V_0$, on
Im $\chi(\omega)$ in the absence of pumping. The value of $(U_0-V_0)$ is
indicated for each curve. The arrows (from left to right) indicate the band
edges for the curves $(4-1.5)$, $(4-2.0)$,
and $(4-2.5)$ respectively. Their positions shift
linearly with $V_0$. On the contrary, the band edge is not affected by
varying $U_0$ (cannot be seen easily from the figure). The energy is in
units of $4\delta t$.
}
\end{figure}

This work is based on the simplified model of an infinite and rigid chain 
and seems to be applicable to other conjugated polymers with
exciton levels as well, such as PPV and polysilane.\cite{PPV1,PPV2}
At the end of this section, we present a calculation for PPV. Its 
optical nonlinearity (for a perfect crystalline sample) is 
found to be of the same order as PDA's. The following parameters 
obtained from the spin density profile of nonlinear excitation are 
used:\cite{Shimoi} $t_0=2.02$ eV, $U_0=2.5t_0$, $V_0=1.3t_0$. The 
displacement of ions due to the double bond alternation is about $\pm 
0.055 {\rm \AA}$. This gives $4 \delta t=1.15$ eV because the lattice 
stiffness for PPV is 5.23 eV/{\rm \AA}.\cite{Shimoi}
From the absorption spectrum being calculated, we estimate the exciton 
binding energy to be 0.6 eV, while earlier calculations range from  
0.4 eV\cite{PPV1} to 0.8 eV.\cite{Shimoi} The exciton radius in PPV is 
about 50 ${\rm \AA}$, which spans 8 unit cells. The inter-chain distance 
for PPV is about 4 {\rm \AA},\cite{Conwell} the absorption coefficient 
for PPV at 400 nm is $2.3\times 10^5$/cm at room temperature, and the 
total exciton lifetime for PPV film is 0.32 ns.\cite{Kepler} From these 
data and the change of refraction index being calculated, we find that the 
optical Kerr coefficient for PPV to be $8\times 10^{-8} {\rm cm}^2/{\rm 
W}$. This is of the same order as the value for PDA. 

However, in actucal practice, PPV is rarely 
used in the study of optical nonlinearity, whether it is resonant or 
off-resonant. This may be related to engineering or chemical problems in 
growing high quality single crystals. These aspects are beyond the 
scope of this paper. Even if a high quality single crystal for PPV 
can be obtained, the inter-chain distance will be much smaller in PPV (3 
to 4 ${\rm \AA}$, comparing to 10 ${\rm \AA}$ for PDA). 
Under this circumstance, the effect of inter-chain coupling probably will 
invalidate our result presented here. Another complication 
for PPV is that, for the intensity of $10^7\ {\rm W/cm^2}$, 
we probably have entered the regime where exciton-exciton annihilation 
is significant.\cite{Kepler} Also, because of the high concentration the 
possibility of excited-state absorption increases. Thus, actual pumping 
efficiency should be lower than that reported here. All of these reasons 
will more or less make our calculation futile for PPV. 

On the other hand, exciton-exciton annihilation may not 
play a significant role in PDA below pump intensity $10^8\  
{\rm W/cm^2}$. A recent paper by Schmid showed that the susceptibility of PDA
does not deviate from a pure $\chi^{(3)}$ behavior until the peak
intensity $10^8\ {\rm W/cm^2}$.\cite{Schmid} This seems to
indicate that, neither exciton-exciton annihilation, nor excite-state 
absorption, is appreciable within the range of our consideration.
One possible reason for the higher threshhold of exciton-exciton 
annihilation is that the exciton radius in PDA (about 
10 ${\rm \AA}$) is much smaller than that in PPV (about 50 ${\rm \AA}$) 
and has higher saturation density. This explains to some extent why the 
same formalism works so well for PDA, but not for PPV. This is 
also supported by the fact that simple estimates on the 
exciton density without such annihilation
effect has been quite consistent with the experiments\cite{Green1,Green2}.

\section{Conclusion}

Resonant optical nonlinearity for conjugated polymers can be understood
using the simple picture of PSF, in which the probe 
signal is reduced because the phase space for final states has been occupied
by the excitons. This model provides only an order of magnitude 
estimate and fails to produce more details such as the probe response over 
the whole spectral range (eg. the band edge absorption), the 
position of the exciton peak, the effect of temperature and Coulomb 
interaction... etc. The present study is able to access the effects of 
various microscopic parameters by using the EOM method. We produced the 
electric susceptibility $\chi(\omega)$ that contains the information 
about the positions and oscillator strength of exciton level and 
conduction band edge. By varying the exciton populations, we can observe 
the change of the resonant oscillator strength and the trend is 
consistent with the PSF model (Eq.(\ref{PSF})). We also calculated the 
optical Kerr coefficient $n_2$ and the value obtained $1.7\times 10^{-8}\ 
{\rm cm}^2/{\rm W}$ agrees well with observations.

It has to be reminded that several complications in a real polymer system 
have been left out to simplify the discussion. 
We have used a rigid and infinite polymer chain while in actual
experiments it is finite and maybe flexible. The phonon degrees of freedom 
will contribute to extra features in the absorption curve such as the 
phonon side-bands.\cite{Bishop} We have also used the quasi-equilibrium 
condition for the exciton gas. In future research, the dynamical 
evolution of exciton 
density can be included by coupling the equation of motion to the 
rate equation of the electron population. Finally, the present theory
has to be modified at high exciton density when exciton-exciton 
interaction plays a more important role and the RPA is no longer valid. This 
may lead to exciton-exciton
annihilation, formation of bi-excitons, and even the 
existence of a gain threshold beyond which lasing can happen.\cite{Yan} 

\acknowledgments
M.~C.~C. wishes to thank T.~M.~Hong and M.~F.~Yang for many
helpful discussions. This work is supported by the National 
Science Council of Taiwan under contract Nos. NSC86-2112-M-007-026 and 
NSC86-2112-M-009-001. The support from the National Center of 
High-performance Computing of Taiwan is also acknowledged.  

\appendix
\section{}
Based on the symmetry of inversion, we can show that the 
terms that are quadratic in $n_{c,vk\si}$ in 
Eqs.~(\ref{main0}) do not affect the dynamics of the total 
polarization. First, the inversion properties of the Bloch states are,
\bea
\Psi_{ck}(-r)&=&-\Psi_{c-k}(r),\cr
\Psi_{vk}(-r)&=&\Psi_{v-k}(r).
\eea
This leads to the following identities:
\be
\mu_{vc}(-k)=\mu_{vc}(k),
\ee
and
\bea
{\cal V}_{\lambda_1 \lambda_2 \lambda_3 \lambda_4}^{-k-k'}
&=& {\cal V}_{\lambda_1 \lambda_2 \lambda_3 \lambda_4}^{kk'}
\pi_{\lambda_1}\pi_{\lambda_2}\pi_{\lambda_3}\pi_{\lambda_4},\cr
\tilde{\cal V}_{\lambda_1 \lambda_2 \lambda_3 \lambda_4}^{-k-k'}
&=& \tilde{\cal V}_{\lambda_1 \lambda_2 \lambda_3 \lambda_4}^{kk'}
\pi_{\lambda_1}\pi_{\lambda_2}\pi_{\lambda_3}\pi_{\lambda_4},
\eea
where $\pi_c=-1$, $\pi_v=1$.
It is clear that ${\cal V}_{\lambda_1 \lambda_2 \lambda_3 \lambda_4}^{kk'}$
changes sign when the number of $c$ indices does not equal to
the number of $v$ indices. 

Second, combining the exciton wave function, $\phi_0^{s,t}(k)=( 
2r_0^{s,t}/\pi )^{1/2}/\{1+\left[(k\pm\pi/2)r_0^{s,t}\right]^2\}$, and 
the Bose-Einstein distribution function, $g_{nK}^{s,t}=(\exp\{\beta 
[\epsilon_n^{s,t}(K)-\mu_{s,t}]\}-1)^{-1}$
(see Sec.~III B), and using the fact that $\Psi_k(\br) = 
\Psi_{k\pm \pi} (\br)$, it is not difficult to see from Eq.~(3.8) that
\be
n_{c,v-k\si}=n_{c,vk\si}.
\ee

Because of the symmetries in Eqs.~(A2), (A3), and (A4), the 
equation for 
$p_{-k}$ has the same form as the equation for $p_{k}$ (see 
Eq.~(\ref{main0}),
with $k$ and $k'$ being replaced by $-k$ and $-k'$) except that the signs 
of the terms quadratic in $n_{c,vk\si}$ are changed. Consequently, 
they do not contribute to the total polarization $\langle P\rangle$, in 
which $p_k$ and
$p_{-k}$ appear through the combination of $\mu_{vc}(k) ( p_k + p_{-k} )$ 
only. Notice that the conclusion may not be valid if the electron population
is not in thermal equilibrium.

\end{multicols}

\begin{references}
\bibitem{U_0} 
Actually, the expressions of ${\cal 
V}^{kk'}_{\lambda_1\lambda_2\lambda_3\lambda_4}$ 
and $\tilde{\cal V}^{kk'}_{\lambda_1\lambda_2\lambda_3\lambda_4}$ in 
Eq.~(\ref{V12}) are the same as those in Eqs.~(8) and (9) of Abe's paper 
for optically active excitons that have zero total momenta.
\bibitem{weak} This choice is the same as in Ref.~15. The 
susceptibility being calculated is not sensitive to 
this parameter. For example, if $t_0$ is increased by $50\%$, 
the height of an absorption peak is changed by
less than 10$\%$ and the shift of peak position is insignificant.
This is reasonable since dominant contribution to the exciton formation
comes from states near the inner band edges.
\end{references}
\end{document}